\documentclass[10pt,journal,twocolumn]{IEEEtran}
\IEEEoverridecommandlockouts
\usepackage{algpseudocode}
\usepackage{algorithm}
\usepackage{bm}
\usepackage{longtable}
\usepackage{tabularx}
\usepackage{amsfonts}
\usepackage{amssymb}
\usepackage{lipsum}
\usepackage{amscd}
\usepackage{graphics}
\usepackage[dvips]{graphicx}
\usepackage{epsfig}
\usepackage{subfigure}
\usepackage[cmex10]{amsmath}
\usepackage{array}
\usepackage{eqparbox}
\usepackage{flushend}
\usepackage{hyperref}
\usepackage{fancyhdr}
\usepackage{titling}
\usepackage{enumerate}

\DeclareMathAlphabet{\mathbsf}{OT1}{cmss}{bx}{n}
\DeclareMathAlphabet{\mathssf}{OT1}{cmss}{m}{sl}
\DeclareMathAlphabet{\mathcsf}{OT1}{cmss}{sbc}{n}

\newcommand{\ie}{{\em i.e.}}
\newcommand{\etc}{{\em etc}}

\newcommand{\iid}{i.i.d.}

\newcommand{\secref}[1]{Section~\ref{#1}}

\newcommand{\tabref}[1]{Table~\ref{#1}}

\makeatletter
\def\blfootnote{\xdef\@thefnmark{}\@footnotetext}
\makeatother

\newcommand{\qed}{\nobreak \ifvmode \relax \else
      \ifdim\lastskip<1.5em \hskip-\lastskip
      \hskip1.5em plus0em minus0.5em \fi \nobreak
      \vrule height0.75em width0.5em depth0.25em\fi}

\def\BibTeX{{\rm B\kern-.05em{\sc i\kern-.025em b}\kern-.08em
    T\kern-.1667em\lower.7ex\hbox{E}\kern-.125emX}}

\usepackage{newlfont}

\begin{document}
\title{Impact of System State Dynamics on PMU Placement in the Electric Power Grid}
\author{\IEEEauthorblockN{K. G. Nagananda}\thanks{The author is with the Department of Electrical Communication Engineering, Indian Institute of Science, Bangalore 560012, INDIA, E-mail: \texttt{kgnagananda@alum.lehigh.edu}}
}

\pagenumbering{gobble}
\date{}
\maketitle

\begin{abstract}
The goal of this paper is to study the impact of the dynamic nature of bus voltage magnitudes and phase angles, which constitute the state of the power system, on the phasor measurement unit (PMU) placement problem. To facilitate this study, the placement problem is addressed from the perspective of the electrical structure which, unlike existing work on PMU placement, accounts for the sensitivity between power injections and nodal phase angle differences between various buses in the power network. A linear dynamic model captures the time evolution of system states, and a simple procedure is devised to estimate the state transition function at each time instant. The placement problem is formulated as a series (time steps) of binary integer programs, with the goal to obtain the minimum number of PMUs at each time step for complete network observability in the absence of zero injection measurements. Experiments are conducted on several standard IEEE test bus systems. The main thesis of this study is that, owing to the dynamic nature of the system states, for optimal power system operation the best one could do is to install a PMU on each bus of the given network, though it is undesirable from an economic standpoint.
\end{abstract}


\section{Introduction}\label{sec:introduction}
In the classic setting, the problem of placing phasor measurement units (PMUs) within an electric power system is divided into two parts: (1) obtaining the optimal or minimum number of PMUs; and (2) finding the optimal locations to install these PMUs on the grid. The problem is cast in the mathematical programming framework with the goal to have either complete or incomplete network observability either in the presence/absence of  zero injection measurements. Zero injection measurements are present when the power system has nodes without generation or load. This subject received considerable attention and is well reported in the literature \cite{Nuqui2005} - \nocite{Gou2008}\nocite{Gou2008a}\nocite{Baldwin1993}\nocite{Milosevic2003}\nocite{Zhang2010}\nocite{Xu2004}\nocite{Azizi2012}
\nocite{Kekatos2012}\nocite{Li2013}\nocite{Fesharaki2013}\cite{Anderson2014}. Most of the existing work employed the topology-based approach to characterize the connectivity between buses or nodes in the power network.

Though the topological structure of the grid was used to address wide-ranging problems in power systems, the complex network \cite{Boccaletti2006} perspective of the electric grid presented a different picture. Research efforts were directed to highlight the drawbacks of the topological structure, which is based solely on degree distribution of nodes in the network. It was reported (see \cite[Section I]{Cotilla-Sanchez2012} and references therein) that electric grids in different geographical locations had different degree distributions, leading to varied topological structures. It was also pointed out that the same grid had different topological structures by carrying out different model-based analyses. This discrepancy was attributed to weak characterization of the electrical connections between network components as provided by the topological structure. Related reports supporting this line of argument were found in \cite{Wu1995} - \nocite{Wu2005}\nocite{Atay2006}\cite{Dorfler2010}, where it was shown that, for many classes of complex networks, characterizing the network structure using degree distribution alone was suboptimal, and had implications on node synchronization and performance of the network.

In order to provide a more pragmatic characterization of the electrical influence between various network components, \cite{Cotilla-Sanchez2012} introduced the notion of the \emph{electrical structure} of the power grid. The measurement of the electrical influence necessitated a metric system, which was devised by deriving the sensitivity matrix and taking its complement to obtain a distance matrix. This metric was termed the \emph{resistance distance}\footnote{The notion of resistance distance was first introduced in \cite{Klein1993}. It denotes the effective resistance between a network of resistors}, and was proved to be a formal distance metric for the power network \cite[Appendix]{Cotilla-Sanchez2012}. The entries of the resistance distance matrix quantified the electrical connectedness between nodes in the network - zero value indicated that two components are perfectly connected, while a large number indicated that the corresponding components have negligible electrical influence on each other. More importantly, it captured the sensitivity between power injections and nodal phase angle differences between various buses in the network. In other words, the electrical structure takes into account the variations in voltage magnitudes and phase angles - commonly referred to as the \emph{state} of the power system - whose dynamics are governed by well known state transition models \cite{Debs1970} - \nocite{Nishiya1982}\nocite{LeitedaSilva1983}\nocite{Morvaj1985}\nocite{Durgaprasad1998}\nocite{BrownDoCouttoFilho2009}
\cite{Hassanzadeh2012}.

On the one hand, we have existing work on PMU placement which utilize the topological structure of the grid, while on the other hand, there are results in the area of complex networks which uncover the drawbacks of the topological structure and at the same time promote the electrical structure accounting for variations in the states of the power system. Furthermore, the system states are time-varying, where the time evolution is described by well known dynamic models. Therefore, a natural question that arises is the following: what implications do the electrical structure with time-varying states have on PMU placement in the power grid? We investigate this question in this paper. To the best of the author's knowledge, this is the first instance of taking into account the impact of the dynamic nature of system states on PMU placement.

The problem is formulated as a series (time steps) of minimization programs, with each time step comprising a binary integer program. The objective is to obtain the minimum number of PMUs at each time step to achieve complete network observability in the absence of zero injection measurements. The connectivity of the network (obtained using the resistance distance matrix) is now a function of time, due to its dependence on the dynamic states of the system. Simulations are conducted for several standard IEEE test bus systems.

At first glance, the results are not surprising: at each time step a new minimum number of PMUs is obtained, and these PMUs are placed on buses corresponding to the position of 1's in the optimal decision vector. However, changing the configuration of the set of PMUs at each time step is infeasible from an engineering point of view. Furthermore, the minimum number of PMUs obtained for two time instants $t_1$ and $t_2$ (say) might be the same, but their configurations could be different depending on the connectivity of nodes at times $t_1$ and $t_2$, respectively. Owing to such infeasibility, our study suggests that for optimal system operation (such state estimation, wide-area monitoring, {\etc}.) the best one could do is to install a PMU on each bus of the grid, though it is undesirable from an economic standpoint.

We begin with an overview of existing literature on PMU placement in \secref{sec:relatedwork}. In \secref{sec:state_dynamics}, we present the dynamic model which describes the time behavior of the power system states. \secref{sec:elec_structure} comprises details of the electrical structure of the grid and the procedure for deriving the resistance distance matrix to be used in the placement problem. The placement problem is formulated in \secref{sec:pmu_placement}, while the experimental results are tabulated in \secref{sec:results}. We conclude the paper in \secref{sec:conclusion}.

\section{Existing work on PMU placement}\label{sec:relatedwork}
In this section, we briefly summarize previously published work on PMU placement. In \cite{Nuqui2005}, \cite{Gou2008}, \cite{Gou2008a}, the PMU placement problem was formulated in the integer linear  programming framework for complete or incomplete network observability with or without zero power injection measurements. In \cite{Baldwin1993}, the minimal PMU set was obtained using a dual search bisecting simulated-annealing algorithm searches for complete network observability, while the location problem was solved using a spanning measurement subgraph. A non-dominated sorting genetic algorithm yielding a Pareto-optimal solution was presented in \cite{Milosevic2003}, where the integer program exhibited nonlinearity in the presence of power injection measurements. Multiple placement solutions was proposed in \cite{Zhang2010}, within the framework of power system dynamic state estimation.

An integer program was formulated in \cite{Xu2004} to include conventional power flow and injection measurements in addition to PMU measurements for maximum network observability. In \cite{Azizi2012}, state estimation using phasor measurements with complete network observability was shown to be linear, where an exhaustive search-based method was devised to solve the placement problem. An estimation-theoretic criteria to optimize PMU placement was considered in \cite{Kekatos2012}, where the problem was solved using a convex relaxation method incorporating system states estimated within a Bayesian framework. In \cite{Li2013}, an information-theoretic measure, namely, mutual information (MI) was employed to address the PMU placement problem, where the objective was to maximize the MI between PMU measurements and power system states to obtain highly ``informative'' PMU configurations. Similar observability and data transmission constraints were considered in \cite{Fesharaki2013} and \cite{Anderson2014}.

However, studies in the aforementioned references did not fully utilize the electrical influence - power injections and phase angle differences - between components in the power grid. Though various case studies were performed, the models considered did not provide a realistic scenario for optimal system operation. Furthermore, the dynamics of the system states were not accounted for, thus motivating the study in this paper.

\section{Power system state dynamics}\label{sec:state_dynamics}
We consider a power system comprising power flow meters, transmission lines and buses. Let us suppose there are $N$ buses in the system. We use $i$ and $j$ to denote the bus indices; $i, j\in \{1,\dots,N\}$. Then, the system state vector to be estimated is of the following form: $\boldsymbol{x} = [\theta_2,\dots,\theta_{N}, V_1,\dots,V_{N}]$, where $\theta_{i}$ and $V_{i}$, denote the phase angle and  magnitude of the voltage at the $i^{\text{th}}$ bus, respectively. The phase angle $\theta_1$ of the reference bus is assumed known.

The power system dynamics is specified by the following general state transition model. Letting $t=1,2,\dots$ to be the time index, the state variable at time $t$ is specified by
\begin{eqnarray}
\boldsymbol{x}_{t} &=& \boldsymbol{f}_{t}(\boldsymbol{x}_{t-1}) + \boldsymbol{w}_{t},\label{eq:state_var}
\end{eqnarray}
where, for every time step $t$, $\boldsymbol{f}_{t}$ denotes the state transition function and $\boldsymbol{w}_{t}$ denotes the state noise vector, which is assumed to be {\iid} Gaussian with known statistics.

\begin{figure*}[!th]
\normalsize
\begin{eqnarray}\label{eq:state_update}
\left[
\begin{array}{c}
\theta_{2,t}\\
\vdots\\
\theta_{N,t}\\
\mathrm{V}_{1,t}\\
\vdots\\
\mathrm{V}_{N,t}
\end{array}
\right] =
\underbrace{\left[
\begin{matrix}
\mathit{f}_{11,t} & \dots & \mathit{f}_{1(2N-1),t}\\
\vdots & \ddots & \vdots\\
\mathit{f}_{(2N-1)1,t} & \dots & \mathit{f}_{(2N-1)(2N-1),t}
\end{matrix}\right]}_{\boldsymbol{F}_{t}}\times
\left[
\begin{array}{c}
\theta_{2,t-1}\\
\vdots\\
\theta_{N,t-1}\\
\mathrm{V}_{1,t-1}\\
\vdots\\
\mathrm{V}_{N,t-1}
\end{array}
\right] +
\left[
\begin{array}{c}
w_{1,t}\\
\vdots \\
w_{(2N-1),t}\\
\end{array}
\right].
\end{eqnarray}
\hrulefill
\end{figure*}

In this section, we present a procedure to compute the state transition function $\boldsymbol{f}_{t}$ at each time step. Relevant references in the existing literature include \cite{Debs1970}, where a unity transition matrix is employed to model the slowly changing system state, while in \cite{Nishiya1982} a constant term is added to the tracking model to improve state forecasting at each time step. In \cite{LeitedaSilva1983}, Holt's exponential smoothing is used to update the parameters of the state transition matrix, which is taken to be diagonal; the updating is performed for every set of new measurements. In \cite{Morvaj1985}, the state transition is modeled using a time-variant diagonal matrix, which is updated based on load forecasting. In \cite{Durgaprasad1998}, the authors employ a unit state transition matrix; however, they account for the effect of adjacent buses on the behavior of the system states and the state transition model is based on the nodal analysis. An overview of existing work in this direction can be found in \cite{BrownDoCouttoFilho2009}.

In this paper, we take into account the effect of adjacent buses on the system states and consider a block diagonal matrix to model the state transition function, a method first devised in \cite{Hassanzadeh2012}. Let us consider the system state vector $\boldsymbol{x}_{t} = [\theta_{2,t},\dots,\theta_{N,t}, \mathrm{V}_{1,t},\dots,\mathrm{V}_{N,t}]$ at time $t$. From \eqref{eq:state_var}, we have

For an observation interval of length $M$, \eqref{eq:state_update} can be compactly written as
\begin{eqnarray}\label{eq:state_updateM}
\boldsymbol{X}_{t}^{([1:M])} = \boldsymbol{F}_{t}\boldsymbol{X}_{t-1}^{([1:M])} + \boldsymbol{W}_{t}^{([1:M])},
\end{eqnarray}
where $\boldsymbol{X}_{t}^{([1:M])} \triangleq [\boldsymbol{x}_{t},\dots,\boldsymbol{x}_{t-M}]$, $\boldsymbol{X}_{t-1}^{([1:M])} \triangleq [\boldsymbol{x}_{t-1},\dots,\boldsymbol{x}_{t-1-M}]$, $\boldsymbol{W}_{t}^{([1:M])} \triangleq [\boldsymbol{w}_{t},\dots,\boldsymbol{w}_{t-M}]$, and $\boldsymbol{F}_{t}$ is the state transition matrix. The dimensions of matrices $\boldsymbol{X}_{t}$, $\boldsymbol{X}_{t-1}$ and $\boldsymbol{W}_{t}$ are $(2N-1)\times M$, while that of $\boldsymbol{F}_{t}$ is $(2N-1)\times (2N-1)$. Note that, for \eqref{eq:state_updateM} to hold, $M \geq 2N-1$. By decoupling the voltage magnitude and phase angles, the state transition matrix can be written as
\begin{eqnarray}\label{eq:statetran_matrix}
\boldsymbol{F}_{t} =
\left[\begin{matrix}
\boldsymbol{A}_{t} & \boldsymbol{0} \\
\boldsymbol{0} & \boldsymbol{B}_{t}
\end{matrix}\right],
\end{eqnarray}
where for every $M$, the sub-matrix $\boldsymbol{A}_{t}$ with dimensions $(N-1)\times(N-1)$ corresponds to the phase angle transitions in $\boldsymbol{X}_{t}^{([1:M])}$, while the sub-matrix $\boldsymbol{B}_{t}$ with dimensions $N\times N$ corresponds to voltage magnitude transitions in $\boldsymbol{X}_{t}^{([1:M])}$. Following the procedure in \cite[Section III]{Hassanzadeh2012}, we obtain the least square estimates of the state transition function corresponding to the phase angles and voltage magnitudes at every time step $t$. From the state transition model \eqref{eq:state_var}, the following expression can be written for the angle transition block $\boldsymbol{A}_{t}$:
\begin{eqnarray}
{x}_{t}(i) &=& \boldsymbol{A}(i)\boldsymbol{x}_{t-1} + {w}_{t}(i),\label{eq:angle_block1}
\end{eqnarray}
where $i=1,\dots,N-1$, ${x}_{t}(i)$ is the $i^{\text{th}}$ entry of $\boldsymbol{x}_{t}$, $\boldsymbol{A}(i)$ is the $i^{\text{th}}$ row of the angle transition matrix $\boldsymbol{A}_{t}$, $\boldsymbol{x}_{t-1}$ is the column vector of phase-angles at time $t$, and ${w}_{t}(i)$ is the $i^{\text{th}}$ entry of the state noise vector $\boldsymbol{w}_{t}$ at time $t$. \eqref{eq:angle_block1} can be compactly written as
\begin{eqnarray}
{x}_{t}(i) &=& \boldsymbol{x}_{t-1}^{\mathrm{T}}\boldsymbol{A}^{\mathrm{T}}(i) + {w}_{t}(i).\label{eq:angle_block2}
\end{eqnarray}
Given an observation interval $M$, we have
\begin{eqnarray}
\nonumber {x}_{t}(i) &=& \boldsymbol{x}_{t-1}^{\mathrm{T}}\boldsymbol{A}^{\mathrm{T}}(i) + {w}_{t}(i),\\
{x}_{t-1}(i) &=& \boldsymbol{x}_{t-2}^{\mathrm{T}}\boldsymbol{A}^{\mathrm{T}}(i) + {w}_{t-1}(i),\label{eq:angle_block3}\\
\nonumber \vdots && \vdots \\
\nonumber {x}_{t-M}(i) &=& \boldsymbol{x}_{t-1-M}^{\mathrm{T}}\boldsymbol{A}^{\mathrm{T}}(i) + {w}_{t-M}(i),
\end{eqnarray}
which can be written as
\begin{eqnarray}\label{eq:state_updateM}
\boldsymbol{x}_{t}^{([1:M])} = \boldsymbol{x}_{t-1}^{([1:M])}\boldsymbol{A}^{\mathrm{T}}(i) + \boldsymbol{w}_{t}^{([1:M])}\label{eq:angle_block4},
\end{eqnarray}
where the dimensions of $\boldsymbol{x}_{t}$, $\boldsymbol{x}_{t-1}$, $\boldsymbol{A}(i)$ and $\boldsymbol{w}_{t}$ are $M\times 1$, $M\times (N-1)$, $1\times (N-1)$ and $M\times 1$, respectively. $\boldsymbol{A}(i)$ is updated once every new measurements are available, and is calculated using least squares method as follows:
\begin{eqnarray}
\hat{\boldsymbol{A}}(i) = \left(\boldsymbol{x}_{t-1}^{([1:M]),\mathrm{T}}\boldsymbol{x}_{t-1}^{([1:M])}\right)^{-1}
\left(\boldsymbol{x}_{t-1}^{([1:M]),\mathrm{T}}\boldsymbol{x}_{t}\right). \label{eq:angle_block5}
\end{eqnarray}
Each row of the phase angle transition matrix is obtained using \eqref{eq:angle_block5}. The same procedure is used to update the sub-matrix $\boldsymbol{B}_t$, corresponding to voltage magnitudes.

\section{Electrical structure of the power network}\label{sec:elec_structure}
In this section, we review the electrical structure of the power network. The concept of resistance distance is introduced, and the procedure to derive the binary connectivity matrix of a given power grid is presented. The exposition presented here follows from \cite[Section III]{Cotilla-Sanchez2012}, with minor differences in the derivation of the distance matrix.

The sensitivity between power injections and nodal phase angles differences can be utilized to characterize the electrical influence between network components. The electrical structure of the power network can then be understood by measuring the amount of electrical influence that one component has on another in the network. The measurement of this electrical influence necessitates a metric system. Mathematically, this can be accomplished by first deriving the sensitivity matrix, which can be obtained by standard methods. The complement of the sensitivity matrix is called the distance matrix, whose entries quantify the electrical influence that each component has on the other - zero value indicates that two components are perfectly connected, while a large number indicates that the corresponding components have negligible electrical influence on each other. This electrical distance was proved to be a formal distance metric, and was employed to address various problems in power systems.

Another method to measure the electrical influence between network components is to derive the resistance distance \cite{Klein1993}, which is the effective resistance between points in a network of resistors. Consider a network with $N$ nodes, described by the conductance matrix $\bm{G}$. Let $V_j$ and $g_{ij}$ denote the voltage magnitude at node $j$ and the conductance between nodes $i$ and $j$, respectively. The current injection at node $i$ is then given by
\begin{eqnarray}\label{eq:current_injection}
I_i = \sum_{j=1}^{N}g_{ij}V_j.
\end{eqnarray}
$\bm{G}$ acts as a Laplacian matrix to the network, provided there are no connections to the ground, {\ie}, if $\bm{G}$ has rank $N-1$. The singularity of $\bm{G}$ can be overcome by letting a node $r$ have $V_r = 0$. The conductance matrix associated with the remaining $N-1$ nodes is full-rank, and thus we have
\begin{eqnarray}\label{eq:nonreferencenodes}
\bm{V}_k = \bm{G}^{-1}_{kk}\bm{I}_k, k \neq r.
\end{eqnarray}
Let the diagonal elements of $\bm{G}^{-1}_{kk}$ be denoted $g^{-1}_{kk}$, $\forall k$, indicating the change in voltage due to current injection at node $k$ which is grounded at node $r$. The voltage difference between a pair of nodes $(i,j)$, $i\neq j\neq r$, is computed as follows:
\begin{eqnarray}\label{eq:voltage_difference}
e(i,j) = g^{-1}_{ii} + g^{-1}_{jj} - g^{-1}_{ij} - g^{-1}_{ji},
\end{eqnarray}
indicating the change in voltage due to injection of $1$ Ampere of current at node $i$ which is withdrawn at node $j$. $e(i,j)$ is called the resistance distance between nodes $i$ and $j$, and describes the sensitivity between current injections and voltage differences. In matrix form, letting $\boldsymbol{\Gamma} \triangleq \text{diag}(\bm{G}^{-1}_{kk})$, we have $\forall k \neq r$
\begin{eqnarray}\label{eq:matrix_form}
\bm{E}_{kk} &=& \boldsymbol{1}\boldsymbol{\Gamma}^{\mathrm{T}} + \boldsymbol{\Gamma}\boldsymbol{1}^{\mathrm{T}} - \bm{G}^{-1}_{kk} - \left[\bm{G}^{-1}_{kk}\right]^{\mathrm{T}},\\
\bm{E}_{rk} &=& \boldsymbol{\Gamma}^{\mathrm{T}},\\
\bm{E}_{kr} &=& \boldsymbol{\Gamma}.
\end{eqnarray}
The resistance distance matrix $\bm{E}$ possesses the properties of a metric space \cite{Klein1993}.

To derive the sensitivities between power injections and phase angles, we start with the upper triangular part of the Jacobian matrix $\bm{J}$ obtained from the power flow analysis. Note that, $\bm{J}$ is a function of $\Delta |V|$ and $\Delta \theta$, which denote the sensitivities in power injections and nodal phase angle differences, respectively, between network components. By letting $\bm{G} = \bm{J}$, the resulting distance matrix $\bm{E}$ measures the incremental change in phase angle difference $(\theta_i - \theta_j)$ between two nodes $i$ and $j$, given an incremental average power transaction between those nodes, assuming the voltage magnitudes are held constant. It was proved in \cite[Appendix]{Cotilla-Sanchez2012} that $\bm{E}$, thus defined, satisfies the properties of a distance matrix, as long as all series branch reactance are nonnegative.

For a power grid with $N$ buses, the distance matrix $\bm{E}$ translates into an undirected graph with $N(N-1)$ weighted branches. In order to compare the grid with an undirected network without weights, one has to retain the $N$ buses, but replace the $K$ branches with $K$ smallest entries in the upper or lower triangular part of $\bm{E}$. This results in a graph of size $\{N,K\}$ with edges representing electrical connectivity rather than direct physical connections. The adjacency or binary connectivity matrix $\bm{C}$ of this graph is obtained by setting a threshold, $\tau$, adjusted to produce exactly $M$ branches in the network:
\begin{eqnarray}\label{eq:adjcency_matrix}
\bm{C}:
\begin{cases}
c_{ij} = 1, ~\forall e(i,j) < \tau,\\
c_{ij} = 0, ~\forall e(i,j) \geq \tau
\end{cases}
\end{eqnarray}
Thus, we see that the matrix $\bm{C}$ is a function of the system states ($\theta$ and $V$) at any time instant. In \secref{sec:results}, we will derive the binary connectivity matrix $\bm{C}$ for several standard IEEE test bus systems for use in the PMU placement problem.

\section{PMU placement problem}\label{sec:pmu_placement}
In this section, we formulate the PMU placement problem based on the electrical structure of the power network, incorporating the dynamic nature of the system states. The problem is formulated as a series of binary integer programs, each with one inequality constraint for complete network observability. We do not consider zero injection measurements in this paper.

Let us consider a power network with $N$ buses and $K$ branches. At time $t$, we let $\bm{C}_t$ denote the binary connectivity matrix of dimensions $N\times N$, $\bm{d}_t = [d_{1,t},\dots,d_{N,t}]$ with dimensions $N\times 1$ denote the binary decision variable vector defined as follows:
\begin{eqnarray}\label{eq:binarydecision_vector}
d_{i,t} = \begin{cases}
1, ~\text{if a PMU is installed at bus}~i,\\
0, ~\text{otherwise},
\end{cases}
\end{eqnarray}
where $i=1,\dots,N$, and $\bm{b}$ is a unit vector of dimensions $N\times 1$. The PMU placement problem at every time step $t$ is formulated as follows:
\begin{eqnarray}\label{eq:pmuplacement}
\nonumber \min \sum_{i=1}^{N}d_{i,t}\\
\text{such that}~ \bm{C}_t\bm{d}_t \geq \bm{b}\\
\nonumber d_{i,t} \in \{0,1\},
\end{eqnarray}
where $\bm{C}_t$ is given by \eqref{eq:adjcency_matrix}.

\section{Experimental results}\label{sec:results}
In this section, we present the experimental setup and tabulate the main results of this paper. The following standard IEEE test bus systems were the subjects of study: IEEE-14, IEEE-30, IEEE-39, IEEE-57 and IEEE-118. The time evolution of the system states, governed by \eqref{eq:state_var}, was observed for 100 time units. At each time step, the state transition matrix, given by \eqref{eq:statetran_matrix}, was estimated according to the procedure described in \secref{sec:state_dynamics}. The state transition noise $\boldsymbol{w}_{t}$ was assumed to be {\iid} across time, sampled from a Gaussian distribution with mean zero and variance $\sigma^2$; for the experiments conducted in this paper, we let $\sigma^2 = 1$. The bus and branch data, required to derive the power-flow Jacobian matrices, were obtained using MATPOWER \cite{Zimmerman2011}. The binary integer programming tool of Matlab was used to solve the problem formulated in \eqref{eq:pmuplacement} at each time unit. The goal was to obtain the minimum number of PMUs at every time instant by solving the placement problem based on the electrical structure of the power network for complete network observability without zero injection measurements.

The results are shown in \tabref{tab:results_minnum}, where the minimum number of PMUs for several IEEE test bus systems over an observation interval of 100 time units are tabulated. Owing to the dynamic nature of voltage magnitudes and phase angles, the network connectivity matrix $\bm{C}_t$ evolves with time; the subscript $t$ signifies the time dependence. Therefore, the minimum number of PMUs obtained at each time step may be different from that obtained during the previous time instant. There are also several instances where the minimum number of PMUs remains constant over an observation interval.

However, it is important to recognize that the PMU configurations ({\ie} placement) vary over time, though the minimum number of PMUs does not change. This fact can be seen in \tabref{tab:results_placement}, where we show the placement of PMUs for the IEEE-14 bus network between the 40$^{\text{th}}$ and 45$^{\text{th}}$ time units. As shown in \tabref{tab:results_placement}, though the minimum number ({\ie}, 7) of PMUs remains the same for the given observation interval, the optimal locations vary.

\begin{table*}[t]
\centering
\normalsize \vline
    \begin{tabular}{l@{\hspace{10pt}} *{14}{| c |}}\hline
    \bfseries Time & \multicolumn{14}{|c|}{\bfseries Bus number} \\ \hline
        & 1 & 2 & 3 & 4 & 5 & 6 & 7 & 8 & 9 & 10 & 11 & 12 & 13 & 14 \\ \hline
        40 & 1 &  0 & 1 & 1 & 0 & 1 &  0 & 1 & 1 & 0 & 0 & 1 &  0 & 0\\ \hline
        41 & 0 &  0 & 0 & 1 & 1 & 1 &  1 & 0 & 0 & 0 & 1 & 1 &  0 & 1\\ \hline
        42 & 1 &  1 & 1 & 1 & 0 & 0 &  0 & 1 & 1 & 0 & 0 & 0 &  0 & 1\\ \hline
        43 & 0 &  0 & 1 & 0 & 0 & 0 &  0 & 1 & 1 & 1 & 0 & 1 &  1 & 1\\ \hline
        44 & 1 &  1 & 0 & 0 & 0 & 1 &  1 & 1 & 1 & 0 & 0 & 1 &  0 & 0\\ \hline
        45 & 0 &  0 & 1 & 1 & 1 & 0 &  0 & 1 & 1 & 0 & 0 & 1 &  0 & 0\\ \hline
    \end{tabular}
    \caption{Placement of 7 PMUs for the IEEE-14 bus network between the 40$^{\text{th}}$ and 45$^{\text{th}}$ time units.}
    \label{tab:results_placement}
\end{table*}

The variations are clearly due to the dynamic nature of the system states, which causes the connectivity matrix to vary with time. Therefore, for each time unit, the optimal decision vector $\bm{d}_t$ might be different from that of the previous or next time instant. Therefore, for wide-area system monitoring, state estimation and control of the power grid, the study conducted in this paper suggests to install a PMU on each bus of the given power network.

Future work would involve extensions to the case of incomplete network observability (for definition, see \cite{Nuqui2005}). Other avenues include the study of implications of the results presented in this paper on state estimation and wide-area monitoring and control. Since the proposition is to populate the power network with PMUs, efficient communication architectures to transmit/receive phasor data across the network also demands attention.

\section{Conclusion}\label{sec:conclusion}
We studied the impact of the dynamic nature of states of the power system on the PMU placement problem. We approached the problem from the viewpoint of the electrical structure to account for the sensitivity between power injections and nodal phase angle differences between various buses in the electric grid. The time evolution of the system states was captured by a linear dynamic model and the state transition function was estimated at each time instant. The placement problem was formulated as a series of minimization programs, with the goal to obtain the minimum number of PMUs at each time step for complete network observability in the absence of zero injection measurements. Experiments were conducted on IEEE-14, IEEE-30, IEEE-39, IEEE-57 and IEEE-118 test bus systems. Unlike existing results on this problem, our study suggests to install a PMU on each bus of the given network for optimal system operation, since (i) the minimum number of PMUs to achieve the desired objective varies with time and (ii) the optimal locations vary with time even though the minimum number of PMUs is the same for these time instants.

\bibliographystyle{IEEEtran}
\bibliography{IEEEabrv,powersystems}
\raggedbottom

\onecolumn
{
\begin{longtable}{|c|c|c|c|c|c|}\hline
   Time & IEEE-14 & IEEE-30 & IEEE-39 & IEEE-57 &  IEEE-118 \\ \hline
    1     & 8     & 22    & 30    & 39    & 91 \\ \hline
    2     & 7     & 21    & 26    & 37    & 81 \\ \hline
    3     & 7     & 21    & 26    & 37    & 80 \\ \hline
    4     & 6     & 20    & 28    & 36    & 115 \\ \hline
    5     & 6     & 19    & 27    & 36    & 78 \\ \hline
    6     & 5     & 19    & 28    & 38    & 79 \\ \hline
    7     & 5     & 19    & 27    & 34    & 78 \\ \hline
    8     & 6     & 18    & 26    & 36    & 78 \\ \hline
    9     & 6     & 18    & 25    & 36    & 77 \\ \hline
    10    & 5     & 17    & 24    & 37    & 79 \\ \hline
    11    & 6     & 17    & 24    & 37    & 80 \\\hline
    12    & 6     & 16    & 24    & 38    & 78 \\\hline
    13    & 6     & 17    & 24    & 38    & 79 \\\hline
    14    & 7     & 17    & 23    & 37    & 77 \\\hline
    15    & 7     & 16    & 23    & 37    & 77 \\\hline
    16    & 7     & 16    & 23    & 37    & 78 \\\hline
    17    & 7     & 17    & 23    & 37    & 79 \\\hline
    18    & 7     & 17    & 23    & 37    & 79 \\\hline
    19    & 7     & 17    & 23    & 38    & 77 \\\hline
    20    & 7     & 17    & 23    & 37    & 78 \\\hline
    21    & 7     & 18    & 23    & 36    & 80 \\\hline
    22    & 7     & 18    & 23    & 36    & 80 \\\hline
    23    & 7     & 18    & 23    & 35    & 80 \\\hline
    24    & 7     & 17    & 22    & 35    & 79 \\\hline
    25    & 7     & 18    & 22    & 36    & 78 \\\hline
    26    & 7     & 18    & 23    & 35    & 78 \\\hline
    27    & 7     & 18    & 22    & 35    & 80 \\\hline
    28    & 7     & 18    & 22    & 35    & 80 \\\hline
    29    & 7     & 18    & 22    & 35    & 80 \\\hline
    30    & 7     & 17    & 23    & 36    & 79 \\\hline
    31    & 7     & 17    & 23    & 36    & 80 \\\hline
    32    & 7     & 18    & 22    & 36    & 80 \\\hline
    33    & 7     & 18    & 22    & 35    & 80 \\\hline
    34    & 7     & 18    & 22    & 35    & 79 \\\hline
    35    & 7     & 18    & 22    & 35    & 79 \\\hline
    36    & 7     & 17    & 22    & 35    & 79 \\\hline
    37    & 7     & 17    & 22    & 35    & 80 \\\hline
    38    & 7     & 17    & 22    & 35    & 78 \\\hline
    39    & 7     & 17    & 22    & 36    & 79 \\\hline
    40    & 7     & 17    & 22    & 35    & 80 \\\hline
    41    & 7     & 18    & 22    & 36    & 80 \\\hline
    42    & 7     & 18    & 22    & 35    & 81 \\\hline
    43    & 7     & 18    & 22    & 35    & 82 \\\hline
    44    & 7     & 18    & 22    & 35    & 82 \\\hline
    45    & 7     & 18    & 22    & 35    & 82 \\\hline
    46    & 7     & 18    & 22    & 35    & 82 \\\hline
    47    & 7     & 18    & 22    & 35    & 83 \\\hline
    48    & 7     & 17    & 22    & 35    & 83 \\\hline
    49    & 7     & 18    & 22    & 35    & 83 \\\hline
    50    & 7     & 17    & 22    & 35    & 83 \\\hline
    51    & 7     & 18    & 22    & 35    & 82 \\\hline
    52    & 7     & 18    & 22    & 35    & 82 \\\hline
    53    & 7     & 18    & 22    & 35    & 83 \\\hline
    54    & 7     & 18    & 23    & 35    & 83 \\\hline
    55    & 7     & 18    & 22    & 35    & 83 \\\hline
    56    & 7     & 18    & 22    & 35    & 83 \\\hline
    57    & 7     & 18    & 22    & 35    & 82 \\\hline
    58    & 7     & 18    & 22    & 35    & 82 \\\hline
    59    & 7     & 19    & 22    & 35    & 83 \\\hline
    60    & 7     & 18    & 22    & 35    & 84 \\\hline
    61    & 7     & 18    & 22    & 35    & 83 \\\hline
    62    & 7     & 18    & 22    & 35    & 83 \\\hline
    63    & 7     & 18    & 22    & 35    & 83 \\\hline
    64    & 7     & 19    & 22    & 35    & 83 \\\hline
    65    & 7     & 18    & 22    & 35    & 83 \\\hline
    66    & 7     & 19    & 22    & 35    & 83 \\\hline
    67    & 7     & 18    & 21    & 35    & 82 \\\hline
    68    & 7     & 18    & 22    & 35    & 82 \\\hline
    69    & 7     & 18    & 22    & 35    & 82 \\\hline
    70    & 7     & 19    & 22    & 35    & 81 \\\hline
    71    & 7     & 18    & 22    & 35    & 82 \\\hline
    72    & 7     & 19    & 22    & 35    & 83 \\\hline
    73    & 7     & 19    & 22    & 35    & 82 \\\hline
    74    & 7     & 19    & 22    & 35    & 81 \\\hline
    75    & 7     & 19    & 22    & 35    & 82 \\\hline
    76    & 7     & 18    & 21    & 35    & 81 \\\hline
    77    & 7     & 18    & 22    & 35    & 82 \\\hline
    78    & 7     & 19    & 22    & 35    & 82 \\\hline
    79    & 7     & 19    & 22    & 35    & 82 \\\hline
    80    & 7     & 19    & 21    & 35    & 83 \\\hline
    81    & 7     & 19    & 21    & 35    & 82 \\\hline
    82    & 7     & 19    & 22    & 35    & 83 \\\hline
    83    & 7     & 19    & 22    & 35    & 83 \\\hline
    84    & 7     & 19    & 21    & 35    & 82 \\\hline
    85    & 7     & 19    & 21    & 35    & 83 \\\hline
    86    & 7     & 18    & 22    & 35    & 82 \\\hline
    87    & 7     & 18    & 22    & 35    & 83 \\\hline
    88    & 7     & 17    & 22    & 35    & 83 \\\hline
    89    & 7     & 19    & 22    & 35    & 83 \\\hline
    90    & 7     & 19    & 22    & 35    & 83 \\\hline
    91    & 7     & 19    & 23    & 35    & 81 \\\hline
    92    & 7     & 18    & 23    & 35    & 82 \\\hline
    93    & 7     & 18    & 22    & 35    & 81 \\\hline
    94    & 7     & 18    & 22    & 35    & 81 \\\hline
    95    & 7     & 18    & 22    & 35    & 82 \\\hline
    96    & 7     & 18    & 22    & 35    & 83 \\\hline
    97    & 7     & 18    & 22    & 35    & 83 \\\hline
    98    & 7     & 18    & 23    & 35    & 82 \\\hline
    99    & 7     & 18    & 22    & 35    & 84 \\\hline
    100   & 7     & 18    & 22    & 35    & 82 \\\hline
    \caption{Minimum number of PMUs for IEEE test bus systems for 100 time steps.}
    \label{tab:results_minnum}
\end{longtable}
}

\end{document}